\documentclass[conferenceproceedings,proceedings,accept,moreauthors,pdftex,10pt,a4paper]{mdpi}  

\firstpage{1} 
\articlenumber{157}
\doinum{10.3390/ecea-4-05010}
\pubvolume{2}
\pubyear{2018}
\copyrightyear{2018}
\history{ Published: 20 November 2017}
\pdfoutput=1

\Title{Causal Classical Physics in Time Symmetric \\Quantum Mechanics $^{\dagger}$}

\Author{Fritz W. Bopp}  
\AuthorNames{Fritz W. Bopp}

\address[1]{Department of Physics, University of Siegen, D-57068 Siegen, Germany; bopp@physik.uni-siegen.de; \mbox{Tel.: +49-271-740-3736}}   

\firstnote{\hangafter=1 \hangindent=1.05em \hspace{-0.82em}Presented at the 4th International Electronic Conference on Entropy and Its Applications, Basel, Switzerland, \mbox{21 November–1 December 2017}. Available online: \url{http://sciforum.net/conference/ecea-4}.} 

\abstract{
The letter submitted is an executive summary
of our previous paper.~To solve the Einstein Podolsky
Rosen ``paradox'' the two boundary quantum
mechanics is taken as self consistent interpretation of quantum dynamics. The difficulty with this interpretation is
to reconcile it with classical physics. To avoid macroscopic backward
causation two ``corresponding transition rules''
are formulated which specify needed properties of macroscopic observations
and manipulations. The~apparent classical causal decision tree requires
to understand the classically unchosen options. They are taken to
occur with an ``incomplete knowledge'' of
the boundary states typically in macroscopic considerations. The precise
boundary conditions with given phases then select the actual measured
path and this selection is mistaken to happen at the time of measurement.
The apparent time direction of the decision tree originates in an
assumed relative proximity to the initial state. Only~the far away
final state allows for classically distinct options to be selected
from. Cosmologically the picture could correspond
to a big bang initial and a hugely extended final state scenario.
It is speculated that it might also hold for a big bang/big crunch
world. If this would be the case the Born probability postulate could
find a natural explanation if we coexist in the expanding and the
correlated CPT conjugate contracting world.}

\begin{document}



\vspace{24pt}

What we know is Quantum Dynamics (called Bra-ket physics) establishing  
stable states and describing matrix elements between an initial state
and a final state or analogous matrix elements in Quantum Field Theory
at a given scale. An unprecedented precision (sometimes with ten digits)
makes the theory most dependable. 

Measurements, however, present a problem. Available concepts are not
part of a well tested theory. As pointed out a century ago by Einstein,
Podolsky and Rosen there is a paradox~\cite{einstein1935can} which
needs to be resolved~%
(It is by no means esoteric, p.e.~\cite{Davidson2017,hellmuth1987delayed}). However this is not a problem of underlying Quantum Dynamics but
of the interface between Quantum Dynamics and Macroscopic Physics.
Hence~there are two ways to resolve it. One can either amend or adjust
Quantum Dynamics or classical macroscopic~concepts.    

Considerable effort was invested in the first option.~The Copenhagen
interpretation~\cite{bohr1935can} of Quantum Mechanics involving
quantum jumps and a limited ontology of waves of fields is reasonable
but ugly. Penrose~\cite{PenroseRoad} is probably correct to speculate
that the Hidden Variable~\cite{bell1964einstein,Hooft:2016Cellular}
and the Guiding Fields~\cite{broglie1927structure,bohm1957discussion,durr2009bohmian}
approaches do not work.~New experiments involving quantum interferences
under gravitation~\cite{Schlippert:2014xla} seem to exclude decoherence
by gravitation~\cite{PenroseRoad}. Deterministic approaches like
the ``Cellular Automaton'' advocated by~'t~Hooft~\cite{Hooft:2016Cellular}
somehow lack motivation as they also lead to similar complications
as the ones discussed below.    

We here search for a solution in the second approach.~The paradigm~\cite{Bopp2017}
is to take the wave functions as real ontological objects following
well established quantum dynamics and to consider classical physics
just as effectively valid.~Aspects of the causal macroscopic picture---we accepted a long time ago as well known---might actually not be
true on a fundamental level. The~huge body of observations could just
be a good approximation somehow reflecting our particular cosmological~situation. 

We do not see a way to describe the multi-faced evolution of the cosmos
in a meaningful mathematical even approximate framework. The picture
presented below is therefore largely limited to intuitive arguments.
It develops needed rules and somewhat daring concepts of required
mechanisms envisioned to recapture our macroscopic concepts.

With real ontological wave functions the Einstein Podolsky Rosen paradox
means instantaneous action over a large distance. Contrary to lore
it does not violate the essence of special relativity which just prescribes
boost transformations~%
(Our hope is that following't~Hooft~\cite{Hooft:2016itl} that general
relativity can be formulated in a unitary way even around singular
structures like black holes and that following Donoghue~\cite{Donoghue:2017vvl}
in absense of of quantum jumps quantum field theory can be implemented
in general relativity involving integrals over products of fields.~That coexisting paths may lead to different final gravitational field
configurations presents no problem for the argument given below). However, using~relativity instantaneous action means backward causation
in a different Lorentz system~\cite{Argaman2010Bell} which is widely
considered unacceptable.    

However it is not as bad as it looks. Quantum dynamics contains no
time arrow~\cite{sakurai2014modern} and backward causation is not
unacceptable.

The usual interpretation of quantum mechanics involves a fixed initial
and asymmetrically an open final state and quantum jumps. The original
initial state can, say, evolve to a state containing an electron with
sidewards (say right) spin. If a Stern Gerlach like measurement finds
an upward spin the original initial state is then replaced by the
new one in a non unitary, non time symmetric jump:
\[
<\mathrm{initial}\,|\,\mathrm{Measurement}\,|\to<\mathrm{initial}\,|\,\mathrm{Projection}\,|\,\,/\left|<\mathrm{initial}\,|\,\mathrm{Projection\,|}\,\right|
\]

The central point is that in a theory with backward causation the
measurement outcome like the up/down decision can just as well be
done at a later time then fixing the observed earlier observation.
\[
<\mathrm{initial}\,|\, U_{1}\,\mathrm{Projection}\, U_{2}\,|\leq\mathrm{initial}\,|\, U_{1}\, U_{2}\,\mathrm{Projection'}\,|
\]

For this the measurement must involve sufficient witnesses so that
the projected state can be identified later on with a suitable new
Projection'-operator even if many projections are involved. It~allows
to apply a two boundary picture where the decision is made by a final
state encoding all the measurement decisions. As for the jump formalism
the unitary quantum dynamics evolution has to be amended. One writes:
\[
<\mathrm{initial}\,|\, U_{1}\,\mathrm{Projection}\, U_{2}\,|\,\mathrm{final}>/<\mathrm{initial}\,|\, U_{1}\,\, U_{2}\,|\,\mathrm{final}>\,.
\]
to obtain a unit total probability. This two boundary formalism is
well developed by Aharonov and coworkers~\cite{aharonov1964time}
and others. We stress that this is not a new theory which replaces
the old one in a tricky way which would require extensive work presumably
of a group of people. The formalism just uses the well established
Quantum Dynamics extended it in a straight forward way. 
\pagebreak

In contrast to the quantum world macroscopic considerations do not
allow for distinct coexisting path ways. A large number of effective
measurements~\cite{joos2013decoherence} must reduce ambiguities
to allow for a macroscopic description. In the two boundary description
these measurements must stored in the final boundary. This means that
the overlap
\[
<\mathrm{inital}\,|\,\mathrm{final}>\sim0.5^{\mathrm{decisions}}
\]
must be tiny which is possible. As also claimed by Aharonov and Cohen~\cite{aharonov2017twotime}
it is a self consistent, time symmetric interpretation of quantum
theory. 

The difficulty is to understand how the causal classical physics can
arise in such a frame work. To~proceed we introduce \textit{two transition
rules} which prohibit simple backward causation in classical~physics. 

The first one is known as no `post selection' with macroscopic devices.~Consider a single photon state moving forward in a fiber.~It is possible
to split the fiber in two and join it again with a macroscopically
prearranged relative phase. The forward going channel can be prohibited
but this will not affect the initial creation probability. As a consequence
of unitarity other channels (like reflection) have to open up. 

\begin{figure}[H]
\centering
\includegraphics[scale=0.33]{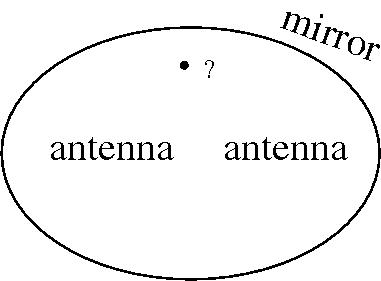}
\caption{Second order interference.}\label{fig1}
\end{figure}

The second rule states that states produced in a macroscopic distance
have a random relative phase. To explain the rule we consider a situation
where it is violated (for similar consideration see~\cite{Alber:2013yea}).
Figure~\ref{fig1} considers two antennas in the focal points of an ellipsoid
mirror. Within the antennae clocked electronics allows to create preselected
situations. With a certain probability one emits a radio frequency
photon at $-\Delta T$ which is than absorbed at the other antenna
at $+\Delta T$. If now both antennae emit at $-\Delta T$ and absorb
at $+\Delta T$ in a symmetric way the probability is not effected.
However if at $t=0$ the mirror gets dark on a point of positive interference
the initial emission probability at $-\Delta T$ is enhanced. This~second order interference effect actually constitutes a violation
of the second rule. The~argument for the rule is that emissions with
a synchronized phases and absorptions not averaging out enhancements
and depletions are extremely rare in macroscopic situations and can
therefore be~ignored. 

In classical physics there is a \textit{causal decision tree}. At
each branching time a decision how the future evolves is made. The
critical point in the considered framework is to understand the unchosen
options. In a macroscopic consideration the quantum phases are averaged
out. With the resulting ``incomplete macroscopic knowledge''
of the boundary states many path ways can appear if the distance between
the boundaries is sufficiently extended. 

The apparent time direction of the decision tree originates in an
assumed relative proximity to the initial state (big bang). Given
the initial and present macroscopic state in all details there is
even with incomplete macroscopic knowledge only one path possible.
It is not easy for macroscopically different states with lots of witnesses
to reach the same final state. Only the assumed really distant final
state allows for multiple chosen and unchosen options. 

If now the exact boundary conditions with their given phases are implemented
the actually taken path is determined. In a classical consideration
this selection is mistaken to happen at the instant of measurement.
So it appears that present decisions affects the choice of the future
path.

The \textit{expanding universe} is source of the thermodynamic arrow~\cite{zeh2001physical}.
In a closed box all intermediate histories will eventually reach a
final state. However, the hugely extended final state of our cosmos
and significant effectively interaction less regions make it plausible
that all of today’s macroscopic decisions can be encoded there. So
the expansion is also source of the effective time arrow considered
here. Of course this is just a conjecture. Many aspects of cosmology
are not well known and it is not clear that in the limit of a large
final time the final state grows faster then needed for the decision
tree. 

In a symmetric scenario with a big bang and a big crunch~\cite{gell1994time,craig1996observation}
it might be enough to have an extremely extended intermediate state
at the turning point. If the initial and final state are identical
any selection collapses as all matching paths contribute and no macroscopic
aspects appear. If they are almost orthogonal both forward and backward
evolutions will produce two distinct extremely entangled intermediate
states. It is conceivable that considering the extreme extend of the
intermediate state the entanglement rarely matches and that effectively
only one intermediate state contributes. 

The unfixed final state opens an amusing possibility. We consider
the situation with an electron wave the time $t$ in the forward moving
world ($t<\frac{1}{2}T_{crunch}$) with spin in the rightward direction
at and an identical one at $T_{\mathrm{crunch}}-t$ in the opposite
moving one. A component $\left\langle \mathrm{rightward}\,|\,\mathrm{upward}\right\rangle $
leads to an upward intermediate state. We assume this state then to
be sufficiently traced in witnesses. The~component which reaches the
same intermediate state in the backward moving world has the same
magnitude $\left\langle \mathrm{rightward}\,|\,\mathrm{upward}\right\rangle {}^{CPT}$. Given the witnesses the common final state allows for no mixed contributions.
In both cases the remaining evolution should happen if a spin is chosen
with unity if the normalization for this case is taken into account.
The probability of an upward spin is~therefore 
\[
P(\mathrm{sideward}\to\mathrm{upward})=|\left\langle \mathrm{sideward}\,|\,\mathrm{upward}\right\rangle |^{2}
\]
and the Born rule is no longer a postulate but a consequence of the
concept. 

The seemingly statistical choice is no longer stored in a know-all
final state but in an intricate miss match of both ``initial'' states.
The overlap is 
\begin{eqnarray*}
\left\langle bang\right| & U(T_{i},t)P_{up}\hspace{6.5cm}\\
 & U(t,T_{match})P_{match}U(T_{match},T_{crunch}-t)\\
 & \hspace{4.3cm}P_{up}U(T_{crunch}-t,T_{crunch}) & \left|crunch\right\rangle 
\end{eqnarray*}
for the upward measurement. A corresponding expression holds for the
downward one. We consider now for both cases the central second line
for the considered case of $P_{match}=1$.~Their values are tiny,
say $10^{-huge}$ resp. $10^{-huge'}$. With $50\%$ the value $huge$
is ``würfelt'' (Einstein's term for dice) larger. The~natural statistical
variation is of the order $10^{-\sqrt{huge+huge'}}$. Given its size
this will always lead to an exclusive dominance of one choice as the
normalization needed in the two boundary formalism is only applied
after summation.

In such a scenario objects would exist with their wave function in
the forward moving world and with their conjugate function eons apart
in tidily correlated opposite moving one.

\vspace{6pt}
\acknowledgments{We acknowledge useful comments of Jos\' e M. Isidro and Giacomo D'ariano. }  

\conflictsofinterest{The authors declare no conflict of interest.}


\reftitle{References}

\end{document}